\documentclass[twocolumn]{aastex631}
\usepackage{amssymb, amsmath, amsthm}
\usepackage{soul} 

\newcommand{\peq}{$P_{\rm eq}$}

\shorttitle{High-density Rocky Planets}
\shortauthors{Lin et al.}

\begin{document}

\title{Most High-density Exoplanets Are Unlikely to Be Remnant Giant-planet Cores}

\correspondingauthor{Zifan Lin}
\email{zifanlin@mit.edu}

\author[0000-0003-0525-9647]{Zifan Lin}
\affiliation{Department of Earth, Atmospheric and Planetary Sciences, Massachusetts Institute of Technology, 77 Massachusetts Avenue, Cambridge, MA 02139, USA}

\author[0000-0001-6294-4523]{Saverio Cambioni}
\affiliation{Department of Earth, Atmospheric and Planetary Sciences, Massachusetts Institute of Technology, 77 Massachusetts Avenue, Cambridge, MA 02139, USA}

\author[0000-0002-6892-6948]{Sara Seager}
\affiliation{Department of Earth, Atmospheric and Planetary Sciences, Massachusetts Institute of Technology, 77 Massachusetts Avenue, Cambridge, MA 02139, USA}
\affiliation{Department of Physics, Massachusetts Institute of Technology, 77 Massachusetts Avenue, Cambridge, MA 02139, USA}
\affiliation{Department of Aeronautics and Astronautics, Massachusetts Institute of Technology, 77 Massachusetts Avenue, Cambridge, MA 02139, USA}



\begin{abstract}

Some exoplanets have much higher densities than expected from stellar abundances of planet-forming elements. There are two theories -- metal-rich formation hypothesis and naked core hypothesis -- that explain how formation and evolution can alter the compositions and structures of rocky planets to diverge from their primordial building blocks. Here, we revisit the naked core hypothesis, which states that high-density planets are remnant cores of giant planets that remain in a fossil-compressed state, even after envelope loss. Using a planetary interior model and assuming energy-limited atmospheric escape, we show that 
a large fraction, if not all, of the iron-silicate core of a giant planet is molten during the planet’s early evolution.
Upon envelope loss, molten part of the planets can rapidly rebound due to low viscosity, resulting in a decrease in radius 
by at most 0.06\%, if they had hydrogen/helium envelopes, or by at most 7\%, if they had H$_2$O envelopes, compared to self-compressed counterparts with the same core mass fraction. 
Based on our findings, we reject the hypothesis that all high-density exoplanets are naked cores with Kolmogorov–Smirnov p-value $\ll$ 0.05 for both envelope compositions.
We find that some high-density exoplanets can still possibly be naked cores, but the probabilities are lower than $\sim1/2$ and $\sim1/3$ for the ice giant and gas giant scenario, respectively, in 95\% of the cases. We conclude that most high-density exoplanets are unlikely to be remnant giant-planets cores. 


\end{abstract}



\section{Introduction} \label{sec:intro}
To date, astronomical observations have revealed over 5,700 exoplanets with highly diverse bulk densities, suggesting a large heterogeneity in compositions and internal structures. Among confirmed rocky exoplanets, there exists a subpopulation of 12 exoplanets (Table \ref{tab:planet_params}) whose densities are significantly larger than most exoplanets with similar sizes. 
This high-density exoplanet subpopulation is currently estimated to account for ${\sim}10\text{--}15\%$ of all terrestrial planets \citep{2021SciAdibekyan,unterborn2022nominal, 2024Cambioni_inprep}. This estimate, if correct, implies that high-density exoplanets are relatively common outcomes of planet formation and evolution.

\begin{figure}[tb]
    \centering
    \includegraphics[width=0.95\linewidth]{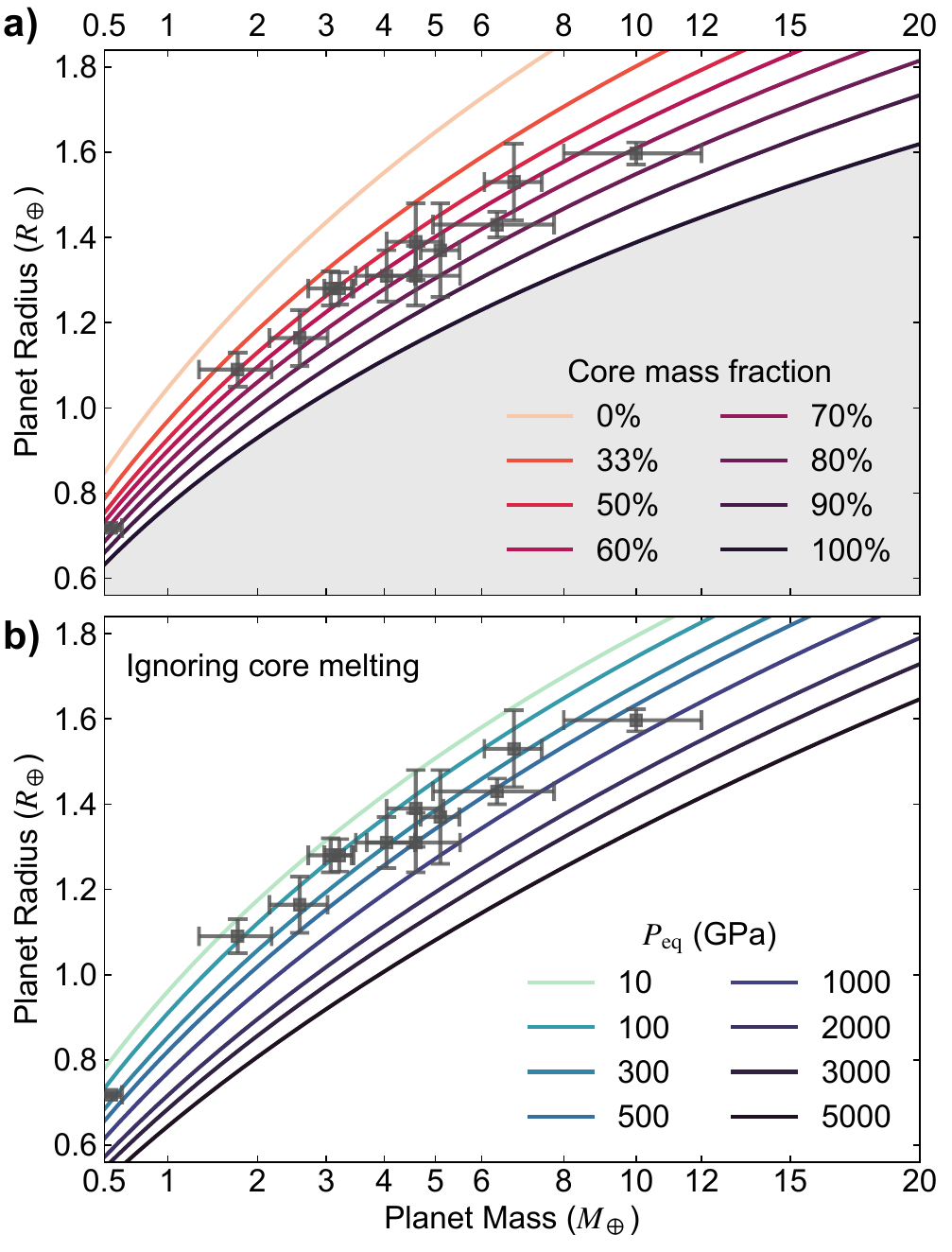}
    \caption{
    \textbf{a)} Mass-radius (M-R) curves of self-compressed rocky planets with CMF ranging from 0\% (pure MgSiO$_3$) to 100\% (pure Fe). \textbf{b)} M-R curves of fossil-compressed naked cores with \peq\ ranging from 10--5000 GPa, \textit{ignoring melting in the iron-silicate core}. 
    The resulting M-R curves considering melting will differ from those shown here.
    In a), the region below the 100\% CMF curve is colored gray to indicate that it is unphysical without external compression. In both panels, measured masses, radii, and their one standard deviation uncertainties of high-density planets are plotted as squares with error bars. 
    Both the iron-rich theory and the naked core hypothesis \citep[as proposed by][]{2014RSPTAMocquet} can explain the masses and radii of high-density planets. However, we revisit the latter herein.}
    \label{fig:mr_combined}
\end{figure}

It is hypothesized that the elemental abundances of planets and their host stars are linked, because they formed from the same reservoir of protoplanetary disk materials \citep[e.g.,][]{2021SciAdibekyan, bonsor_host-star_2021, schulze_probability_2021}. The bulk densities of high-density planets, however, are greater than the 99.7th percentile of planetary densities expected from stellar abundances of planet-forming elements   \citep{2021SciAdibekyan,unterborn2022nominal,2024Cambioni_inprep}. Therefore, there must be formation and evolution mechanisms that can either cause bulk compositions of planets to diverge from protoplanetary disk materials, or alter planetary structures, making some planets denser than others with the same bulk composition.

There are two proposed theories explaining the origin of high-density planets. We review these in turn.

The first theory posits that high-density planets are dense because they have higher iron contents ($\geq50$--60 wt.\%) than other terrestrial planets. To account for their high densities and assuming an internal structure differentiated into a mantle and a core, the core mass fractions (CMF) of these planets are inferred to be $\gtrsim50\%$ (Figure \ref{fig:mr_combined}), many consistent within uncertainties with CMF $\approx70\%$, akin to that measured for planet Mercury by the NASA MESSENGER mission \citep{hauck_ii_curious_2013}. The latter is significantly higher than the CMF of Earth ($\approx32\%$; \citealt{mcdonough_compositional_2003}) and the current estimate for Venus ($\approx32\%$; \citealt{aitta_venus_2012}), which are consistent with the solar iron-to-silicate mass fraction ($33.2\pm1.7\%$; \citealt{2014AJHinkel, 2021SciAdibekyan}). There are two ways to form iron-rich planets. Iron-rich planets may form by preferentially accreting high-density materials near the condensation/sublimation line of refractory elements in the innermost part of protoplanetary disks \citep{2020ApJAguichine,2022AAJohansen}. Alternatively, iron-rich planets may be the remnant cores of differentiated super-Earths that lost their rocky mantles. Mantle-stripping giant impact is the most common explanation of the production of such metallic cores \citep{2009ApJMarcus,2022MNRASReinhardt}. However, recent studies suggest that most super-Earths are too large for their mantle to have been stripped by giant impacts during their formation and evolution \citep{2022ApJScora,2024Cambioni_inprep}.

The second theory explaining the origin of high-density planets hypothesizes that such planets are dense not because of higher iron contents, but because they remain in a \textit{fossil-compressed} higher pressure state than \textit{self-compressed} planets that are compressed only by self-gravity \citep{2014RSPTAMocquet}. In this theory, high-density planets are remnant giant planet cores that once resided under thick volatile envelopes composed of hydrogen and helium (H/He) or H$_2$O. This ``naked core'' hypothesis posits that when giant planets lose their envelopes due to mechanisms such as photoevaporation and tidal stripping, the cores may avoid relaxation and remain in a fossil-compressed state as if the envelope is still exerting pressure, giving rise to their high densities \citep{2014RSPTAMocquet}. This hypothesis has been invoked to explain the origin of some recently discovered dense rocky planets \citep[e.g.,][]{AzevedoSilva_HD_2022, Barros_HD_2022, 2023AJEssack, Guenther_new_2024}.

We cannot distinguish between the iron-rich theory and naked core theory as formulated by \citet{2014RSPTAMocquet}, because both theories match the observed masses and radii of high-density planets (Figure \ref{fig:mr_combined}). An exoplanet consistent with the CMF of Mercury, $\approx70\%$, has the same bulk density as a planet of Earth-like composition that retains the equivalent pressure it had prior to envelope loss, defined as \peq, of several hundreds of GPa (Figure \ref{fig:mr_combined}). Because mass and radius are the primary constraints on the interior composition of exoplanets, additional theoretical examination is needed to differentiate between the two hypotheses.


\begin{table*}
\centering
\begin{tabular}{lcccccl} 
\hline
\hline
Planet Name	&	$R\,(R_\oplus)$			&	$M\,(M_\oplus)$			&	$p_{\rm HD}$\% & $P$ (day) & Age (Gyr) & Reference \\
\hline
GJ 367 b	&		$0.72\pm0.05$	&		$0.55\pm0.08$		&	93  & 0.32 & --- & \cite{lam_gj_2021}  \\
K2-229 b	&		$1.16^{+0.07}_{-0.05}$		&		$2.59\pm0.43$		&	100 & 0.58 & $5.40^{+5.20}_{-3.70}$ & \cite{santerne_earth-sized_2018} \\
GJ 3929 b	&		$1.09\pm0.04$		&		$1.75^{+0.44}_{-0.45}$		&	68	& 2.62 & $7.10^{+4.10}_{-4.90}$ & \cite{beard_gj_2022}  \\
TOI-1468 b	&		$1.28\pm0.04$		&		$3.21\pm0.24$		&	89	& 1.88 & --- & \cite{chaturvedi_toi-1468_2022}  \\
TOI-431 b	&		$1.28\pm0.04$		&		$3.07\pm0.35$		&	77	& 0.49 & --- & \cite{osborn_toi-431hip_2021}  \\
HD 137496 b	&		$1.31^{+0.06}_{-0.05}$		&		$4.04\pm0.55$		&	91	& 1.62 & $8.3\pm0.7$ & \cite{AzevedoSilva_HD_2022}  \\
Kepler-105 c	&		$1.31\pm0.07$		&		$4.60^{+0.92}_{-0.85}$		&	92	& 7.13 & $3.47^{+1.29}_{-1.61}$ & \cite{jontof-hutter_secure_2016, morton_false_2016}  \\
HIP 29442 d	&		$1.37\pm0.11$		&		$5.1\pm0.4$		&	88	& 6.43 & $9.1\pm3.5$ & \cite{damasso_compact_2023, akana_murphy_tess-keck_2023}  \\
L 168-9 b	&		$1.39\pm0.09$		&		$4.60\pm0.56$		&	78	& 1.40 & --- & \cite{astudillo-defru_hot_2020}  \\
Kepler-406 b	&		$1.43\pm0.03$		&	$6.35\pm1.4$		&	94 & 2.43 & $2.14^{+1.56}_{-0.81}$ & \cite{marcy_masses_2014, morton_false_2016}  \\
Kepler-80 d	&		$1.53^{+0.09}_{-0.07}$		&		$6.75^{+0.69}_{-0.51}$		&	81 & 3.07 & $2.88^{+2.73}_{-1.24}$ & \cite{macdonald_dynamical_2016, morton_false_2016}  \\
Kepler-107 c	&		$1.60\pm0.03$		&		$10.0\pm2.0$		&	96 & 4.90 & $4.29^{+0.70}_{-0.56}$ & \cite{bonomo_cold_2023}  \\
\hline
\end{tabular}
\caption{Radius ($R$), mass ($M$), orbital period ($P$), and host star age (if known) of exoplanets with probability of being high-density ($p_{\rm HD}$) greater than 68\%. $p_{\rm HD}$ is defined as the probability that the planet's density meet the high-density criterion from \cite{unterborn2022nominal}, when considering all densities values bootstrapped within M-R measurement uncertainties.}
\label{tab:planet_params}
\end{table*}

\subsection{This Paper}

In this paper, we revisit the naked core hypothesis proposed by \citet{2014RSPTAMocquet}. Specifically,
the current formulation of the theory have three limitations 
that affect its robustness. We review each in turn.

The first limitation 
is that, as \cite{2014RSPTAMocquet} also pointed out, the naked core hypothesis is best applied to massive (30--$100\,M_\oplus$) dense planets, but not the low-mass high-density planets considered here (Table \ref{tab:planet_params}; see also Figure \ref{fig:mr_combined}). Low-mass high-density planets are not massive enough to trigger runaway accretion of H/He. Traditionally, it is assumed that runaway H/He accretion begins when the protoplanet reaches the crossover mass, when the heavy-element mass is roughly equal to the H/He mass \citep[e.g.,][]{stevenson_formation_1982, Bodenheimer_Calculations_1986, helled_mass_2023}. This crossover mass is around $\sim20$--$30\,M_\oplus$, corresponding to a iron-silicate core mass of $\sim10$--$15\,M_\oplus$ (the exact value is model-dependent). We adopt the $\sim10\,M_\oplus$ limit because the most massive high-density planet has a mass of $10.0\pm2.0\,M_\oplus$. Below this limit, planets can accrete some H/He envelopes roughly as massive as the iron-silicate core, but not thick envelopes capable of supplying the hundreds of GPa of pressure needed to compress the naked cores (Figure \ref{fig:mr_combined}).

The second limitation 
is that \cite{2014RSPTAMocquet} considered only compression by H/He, neglecting compression by H$_2$O. However, water is prevalent in the interiors of ice giants such as Uranus and Neptune, which are thought to be dominated by water ice \citep[$\gtrsim70\%$ by mass assuming differentiated layered structure, e.g.,][]{fortney_interior_2010, nettelmann_new_2013}. Such a thick ice layer is expected to exert a pressure of $\gtrsim500$ GPa onto Uranus' and Neptune's Earth-sized iron-silicate cores. This suggests that the progenitors of naked cores could have been Neptune- or sub-Neptune-sized planets. Consistently, the host-star metallicity distribution of ultra-short-period (USP) planets (some of which are high-density, Table \ref{tab:planet_params}) is indistinguishable from that of stars with short-period sub-Neptunes with sizes between 2 and 4 $R_\oplus$. Based on this, \cite{Winn_Absence_2017} concluded that some USP planets could be the solid cores of sub-Neptunes, but they are unlikely to be the cores of evaporated hot Jupiters, which are preferentially found around metal-rich stars.

The third, and most important, limitation 
is that the naked core theory as proposed by \cite{2014RSPTAMocquet} assumes that the iron-silicate core remains solid throughout the envelope's evolutionary history. Thick global magma oceans (MOs), however, are likely to be common on planets with massive H/He or steam envelopes that serve as thermal blankets to keep the interior warm \citep[e.g.,][]{kite_atmosphere_2020}. If a molten magma layer is present, it cannot remain in a fossil-compressed state due to much lower viscosity of melt compared to solid rocks, but will expand as pressure exerted on the core is reduced, lowering bulk planet density.

To address the above three limitations, we use a planetary interior model to simulate the interior structure of naked cores shrouded by both H/He and H$_2$O envelopes and answer the following questions: How much of the core can remain solid throughout the envelope evolution? If a substantial part of the core is molten, how will the final mass-radius (M-R) relation of naked cores be impacted? Throughout this paper, we use the term ``core'' to refer to the iron-silicate inner layers surrounded by H/He or H$_2$O envelope of a planet, as opposed to the conventional definition referring to the innermost iron part of a planet (we call the latter ``iron core''). We use ``atmosphere'' and ``envelope'' interchangeably to refer to the H/He and H$_2$O layers, but note that the inner envelope is not necessarily gaseous under high pressure. 

This paper is structured as follows. Section \ref{sec:methods} describes our planetary interior model. Our results and their implications are presented in Section \ref{sec:results}, and we summarize our conclusions in Section \ref{sec:dis_conclusion}.

\section{Methods} \label{sec:methods}
\subsection{Planetary Interior Model}
We build a planetary interior model that simulates the interior structure of a nonrotating, spherically symmetric planet using three fundamental equations \citep{zapolsky_mass-radius_1969}, namely the mass of a spherical shell
\begin{equation} \label{eq:mass_in_shell}
    \frac{dm(r)}{dr} = 4 \pi r^2 \rho(r),
\end{equation}
hydrostatic equilibrium
\begin{equation} \label{eq:hydro_eq}
    \frac{dP(r)}{dr} = -\frac{Gm(r)\rho(r)}{r^2},
\end{equation}
and the equation of state (EOS)
\begin{equation} \label{eq:eos}
    \rho(r) = f(P(r), T(r)).
\end{equation}
The model iteratively solves for the above equations starting from the planet's center for a given central pressure and integrates outwards with a default step size of 100 m until reaching the desired outer boundary conditions. This formulation is similar to other existing planetary interior codes \citep[e.g.,][]{valencia_internal_2006, seager_massradius_2007, rogers_framework_2010}. 
For naked cores that remain in a fossil-compressed state, the outer boundary condition is met when surface pressure, $P_{\rm surf}$, equals \peq. For self-compressed planets, the outer boundary condition is met when $P_{\rm surf}$ drops below some threshold, which we choose to be 0.1 bar. 

We assume that all simulated planets have a fully differentiated three-layer structure: a Fe core, an MgSiO$_3$ mantle, and a primordial envelope consists of H/He (the ``gas giant'' structure) or H$_2$O (the ``ice giant'' structure).
For Fe core and MgSiO$_3$ mantle, we adopt the isothermal adapted polynomial EOS for iron and silicate \citep{zeng_new_2021}. MgSiO$_3$ is commonly chosen to represent silicate rocks in planetary interior models \citep[e.g.,][]{seager_massradius_2007, wagner_rocky_2012, boujibar_superearth_2020} because its elemental ratio is similar that of the Earth's mantle. The choice of isothermal EOS is justified because thermal expansion leads to negligible density change for iron and silicates, especially under high pressures expected inside naked core progenitors \citep[see e.g.,][]{zapolsky_mass-radius_1969, seager_massradius_2007}. For H$_2$O envelope, we adopt AQUA, a wide pressure-temperature (P-T) range EOS database that incorporates multiple H$_2$O EOSs from the literature and connects H$_2$O phase boundaries using thermodynamically consistent interpolation \citep{haldemann_aqua_2020}. For H/He, we adopt a wide P-T range EOS database \citep{chabrier_new_2019, chabrier_new_2021}, assuming a solar helium mass fraction of $0.275$.

To investigate magma ocean thickness on naked core progenitors, our model further solves for an adiabatic temperature profile. Thermal structure of the iron-silicate core is described by an adiabatic temperature gradient, following \cite{boujibar_superearth_2020}
\begin{equation}\label{eq:boujibar_ad_t}
    \frac{dT_{\rm ad}}{dr} = -\frac{\rho g \gamma T}{K_s},
\end{equation}
where $\gamma$ is the Grüneisen parameter and $K_s$ is the isentropic bulk modulus. For the material thermodynamic constants needed to compute the adiabatic temperature gradient, we adopt post-perovskite values for the silicate mantle and liquid or solid Fe values for the iron core, depending on whether the core is molten (see Table 1 in \citealt{boujibar_superearth_2020}). P-T profile of the H/He or H$_2$O envelope is also computed using an adiabatic gradient, which takes a different form 
\begin{equation}
    \left. \left( \frac{\partial \ln T_{\rm ad}}{\partial \ln P} \right)  \right|_S = \nabla_{\rm ad},
\end{equation}
where $\nabla_{\rm ad}$ is the adiabatic gradient directly supplied by EOS databases \citep{haldemann_aqua_2020, chabrier_new_2019, chabrier_new_2021}.

We thoroughly validate our planetary interior model in Appendix \ref{sec:appendix_validation}.

\subsection{Evolution of Naked Cores}

Here, we propose a structural evolution of naked cores and their progenitors (Figure \ref{fig:flowchart}), describe how this differs from that presented by \citet{2014RSPTAMocquet}, and discuss how we use our planetary interior model to simulate this evolution.

\subsubsection{Progenitors of Naked Cores}

\begin{figure}[t]
    \centering
    \includegraphics[width=\linewidth]{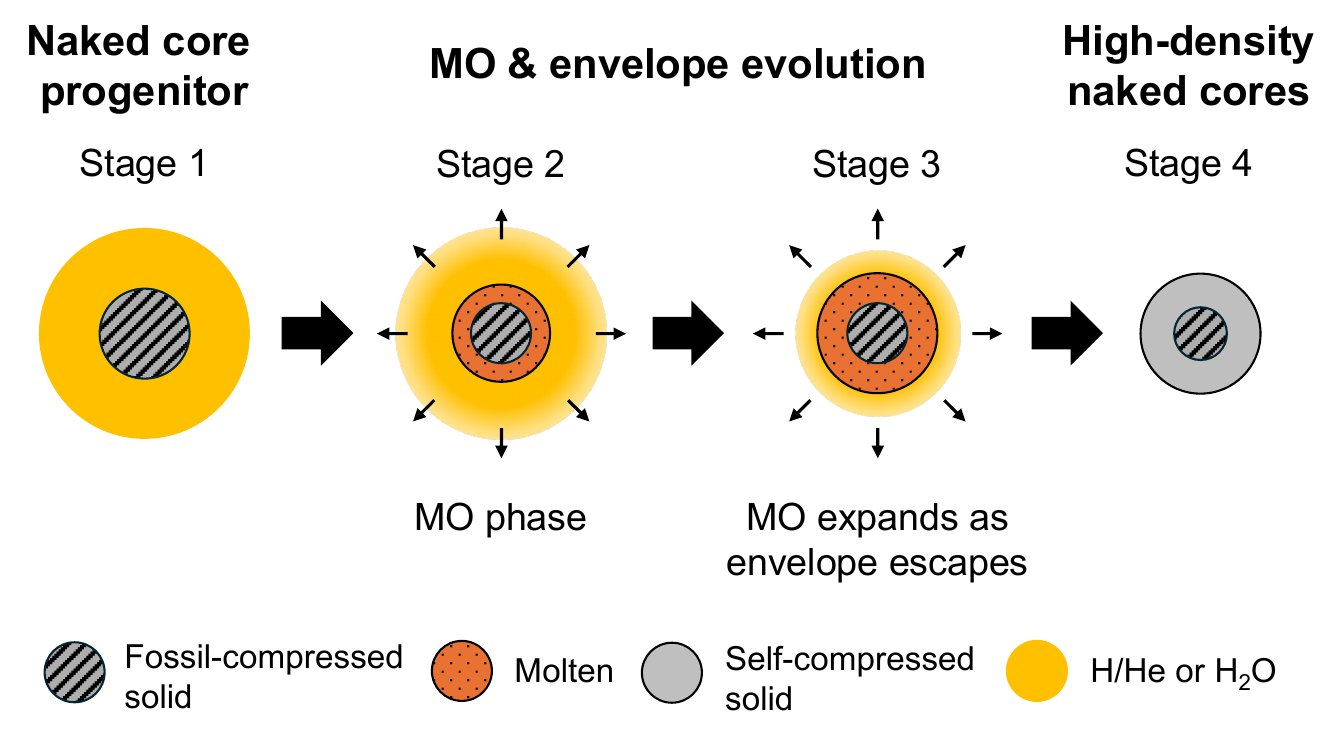}
    \caption{Evolutionary history of naked cores. There are five stages from left to right: (1) The gas giant or ice giant progenitor stage. The progenitor consists of a fossil-compressed iron-silicate core and an outer H/He or H$_2$O envelope that exert a pressure, \peq, on the core. (2) The magma ocean stage, where adiabatic P-T profile is calculated. MO is present wherever $T$ is greater than the melting temperature of iron or silicate. (3) The atmospheric escape stage, where the MO rebounds due to short relaxation timescale of melt, as the envelope undergoes energy-limited escape. (4) The high-density naked core stage, where the envelope is lost completely while the core stays (partially) fossil-compressed.}
    \label{fig:flowchart}
\end{figure}

All naked cores started as fossil-compressed solid cores below H$_2$O or H/He envelopes, as posited by the naked core theory (\citealt{2014RSPTAMocquet}; Stage 1, Figure \ref{fig:flowchart}). 
The iron-silicate portion of the naked core is assumed to have a 33\% CMF, which is the value expected from the galactic average abundance of iron and rock-forming elements \citep{2014AJHinkel, 2024Cambioni_inprep} and close to the CMF of the Earth \citep{mcdonough_compositional_2003}.

Assuming naked cores can indeed remain solid and fossil-compressed after envelope loss, we generate M-R curves in Figure \ref{fig:mr_combined}a by simulating 150 cores with masses, $M_{\rm core}$, spanning 0.5--$20\,M_\oplus$. The mass of the envelope added onto the core is tuned such that desired \peq\ between 10 and 5000 GPa is reached on the core-envelope boundary.

\begin{figure*}
    \centering
    \includegraphics[width=\linewidth]{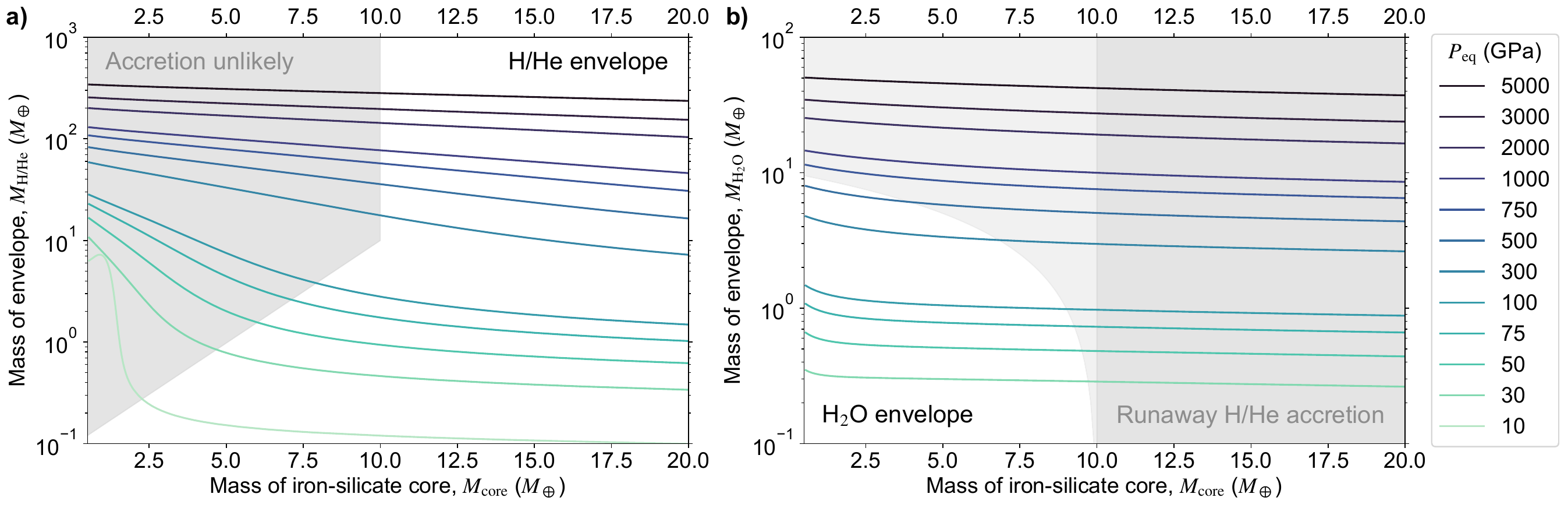}
    \caption{
    Envelope mass of naked core progenitors with \textbf{a)} H/He and \textbf{b)} H$_2$O envelopes.
    We demarcate unphysical or unlikely regions in gray. For gas giants (panel a), only iron-silicate cores $\gtrsim10\,M_\oplus$ are capable of accreting H/He envelopes much more massive than the core itself (Section \ref{sec:intro}).
    For ice giants (panel b), runaway H/He accretion can occur if the combined mass of iron-silicate core and H$_2$O envelope exceeds $10\,M_\oplus$ (light gray region) or if the core has mass $\geq 10\,M_\oplus$ (dark gray region). In both scenarios, the envelope should be H/He-H$_2$O mixture instead of pure H$_2$O.
    Iron-silicate cores below $\sim10\,M_\oplus$ (Table \ref{tab:planet_params}) can only accrete H/He envelopes below $P_{\rm eq}=300$ GPa. Accretion of H$_2$O envelope without significant H/He accretion is limited to $\approx300$ GPa too.
    }
    \label{fig:mcore_menv}
\end{figure*}

Note that according to planet formation models, not all naked core progenitors we synthesize here are physical. Massive envelopes are required to compress iron-silicate cores to the desired \peq. Required envelope mass ranges from a fraction of $M_{\rm core}$ to several hundreds (for gas giants) or tens (for ice giants) of Earth masses (Figure \ref{fig:mcore_menv}), much more massive than the core itself. Naked cores with masses below the crossover mass (equivalent to $\sim10\,M_\oplus$ iron-silicate mass), however, are incapable of accreting and sustaining a H/He envelope much more massive than the core itself \citep[e.g.,][]{stevenson_formation_1982, Bodenheimer_Calculations_1986, helled_mass_2023}. Therefore, for high-density planets detected to date, which have masses $\lesssim 10\,M_\oplus$ (Table \ref{tab:planet_params}), the maximum reachable \peq\ is roughly 300 GPa. For completeness, however, we do not discard unphysical cases with small $M_{\rm core}$ and large \peq\ (gray regions in Figure \ref{fig:mcore_menv}a). 

The thickness of H$_2$O envelope, in contrast, is not limited by core mass. However, as the combined mass of core and H$_2$O envelope exceeds $\sim10\,M_\oplus$, runaway accretion of H/He can be triggered, making a pure H$_2$O envelope unlikely (gray regions in Figure \ref{fig:mcore_menv}b). Due to uncertainty of the crossover mass and for completeness, we do not discard those models either, with the caveat that their thermal structures will approach those models with H/He envelopes if runaway gas accretion is indeed triggered. This will not impact the robustness of upper bounds we place on naked core bulk densities, because H/He envelope maintains the core at a higher temperature than H$_2$O envelope, given the same surface temperature (see Section \ref{sec:results}).


\subsubsection{Magma Ocean and Envelope Evolution}

Here, we reexamine the assumption that naked cores remain fossil-compressed and solid throughout the evolutionary history. We use an adiabatic P-T profile to find MO thickness on an iron-silicate core, as a naked core progenitor gradually loses its atmosphere (Stage 2--3, Figure \ref{fig:flowchart}). MO thickness is solved by comparing the adiabatic P-T profile of the core with melt curves of iron and MgSiO$_3$ (see Section \ref{sec:melt_curves}). Using this comparison, we obtain the molten mass fraction (MMF) of both iron and silicate layers of the naked core.

We simulate atmospheric loss by decreasing \peq. The following \peq\ values are simulated: 5000, 3000, 2000, 1000, 750, 500, 300, 100, 75, 50, 30, and 10 GPa.

We model the envelope evolution of naked core progenitors assuming that energy-limited atmospheric escape is solely responsible for stripping the H/He or H$_2$O envelope \citep[e.g.,][]{luger_extreme_2015, owen_evaporation_2017}. 
We use the expression for the timescale of envelope evaporation presented in \cite{owen_evaporation_2017} to calculate the $T_{\rm eq}$ required to fully evaporate an envelope within a certain timescale, chosen to be 1 Gyr. The detailed formalism is presented in Appendix \ref{sec:appendix_evaporation}.

\subsubsection{Naked Core with Expanded Magma Ocean} \label{sec:mo_viscosity}

MO expands much more efficiently than solid when pressure exerted on it reduces. The relaxation timescale of a viscoelastic Maxwell material is $\tau \propto \eta/K$, where $\eta$ is viscosity and $K$ the bulk modulus. At 150 GPa and 4000 K, $\eta$ of MgSiO$_3$ melt is $\sim1$ Pa s, which can be lower if temperature is higher or if H$_2$O is added to the melt \citep{karki_viscosity_2010}. At the same pressure, $\eta$ of MgSiO$_3$ solid is $\sim10^{23}$ Pa s \citep{stamenkovic_thermal_2011}. The bulk modulus of MgSiO$_3$ melt is $\sim700$ GPa at $P=150$ GPa \citep{spera_structure_2011}, and that of MgSiO$_3$ solid is $\sim800$ GPa at the same pressure \citep{stamenkovic_thermal_2011}. Therefore, relaxation timescale difference between liquid and solid silicates is dominated by the $\sim23$ orders of magnitude difference in viscosity. While it takes $\sim10^9$ years for a fossil-compressed solid core to relax \citep{2014RSPTAMocquet}, the relaxation of a MO is essentially instantaneous.

\subsubsection{The End of Naked Core Evolution}

Depending on the rheological properties of rocky materials, a (partially) fossil-compressed naked core may rebound to a lower pressure state. This relaxation timescale can be on the order of billions of years \citep{2014RSPTAMocquet}. We do not model this stage explicitly because detailed investigation of rheological properties under extreme pressures goes beyond the scope of this study.

\subsection{Iron and Silicate Melt Curves} \label{sec:melt_curves}
To robustly place upper limits on the bulk densities of naked cores, we choose to adopt Fe and MgSiO$_3$ melt curves with the highest temperatures. As such, MMF is minimized and a larger fraction of the planet remains solid throughout the evolution. The choice of melt curve is important because melting of Fe and MgSiO$_3$ at extremely high pressures ($\gtrsim1000$ GPa) is not well constrained by experiments, and theoretical predictions using different approaches can drastically differ.

For MgSiO$_3$, we adopt a melt curve fitted to experimental data by \cite{fei_melting_2021}, which is $\sim$10,000 K hotter at 5000 GPa than the melt curve fitted to another experimental work by \cite{fratanduono_thermodynamic_2018}.

\begin{figure*}
    \centering
    \includegraphics[width=\linewidth]{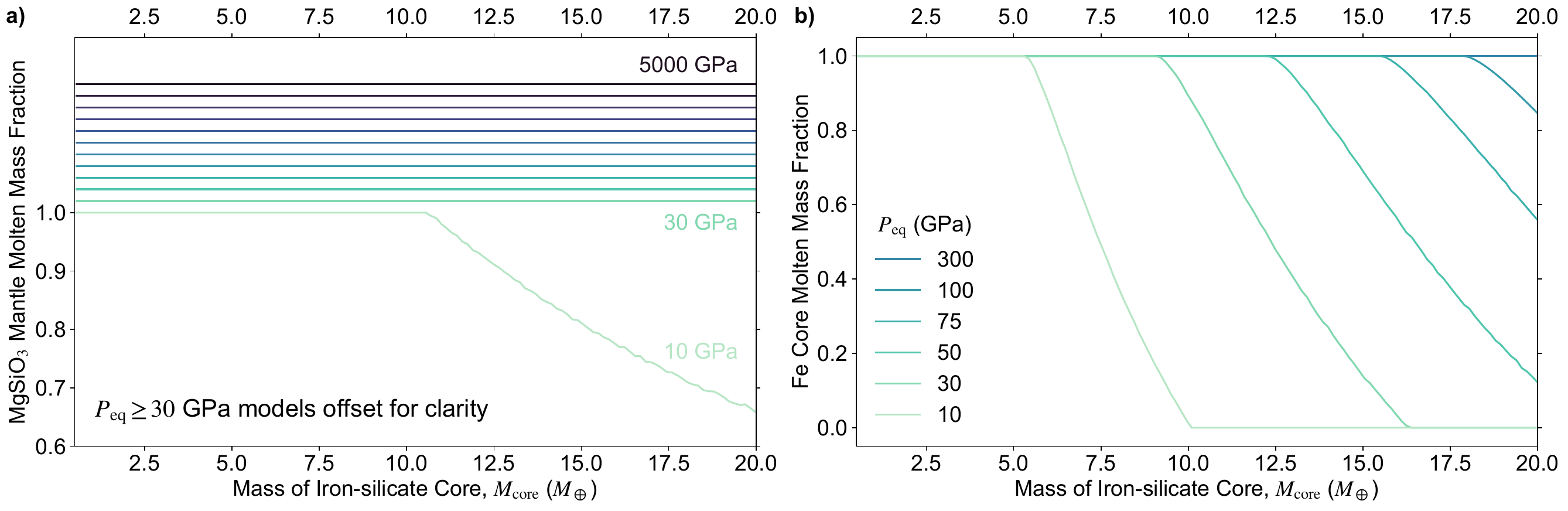}
    \caption{
    Molten mass fractions of \textbf{a)} MgSiO$_3$ mantle and \textbf{b)} iron core for naked cores surrounded by H/He envelopes.
    Curve with darker color shows MMF assuming higher \peq. All mantle MMF equal one, except for the thinnest (\peq\ $=10$ GPa) envelope, suggesting extensive and long-lived MO. The \peq\ $\geq30$ GPa mantle MMF lines are offset by a constant value for clarity. Naked cores with \peq\ $\leq 100$ GPa may have partially solid iron cores, where the mass fraction of solid iron increases with $M_{\rm core}$. When \peq\ $\geq 300$ GPa, all naked cores with H/He envelopes have fully molten iron core. 
    Global melting invalidates the naked core hypothesis (see text).}
    \label{fig:MMF_hhe}
\end{figure*}

For Fe melt curve, we adopt the recent ab initio molecular dynamics simulation presented by \cite{gonzalez-cataldo_ab_2023}, which found the melting temperature of iron by matching the Gibbs free energy for the solid and liquid phases. Melting temperatures reported by \cite{gonzalez-cataldo_ab_2023} are hotter than those extrapolated from earlier experiments by $\sim5000$ K at 5000 GPa, making them the right tool for placing upper limits on naked core bulk densities.

\begin{figure}
    \centering
    \includegraphics[width=\linewidth]{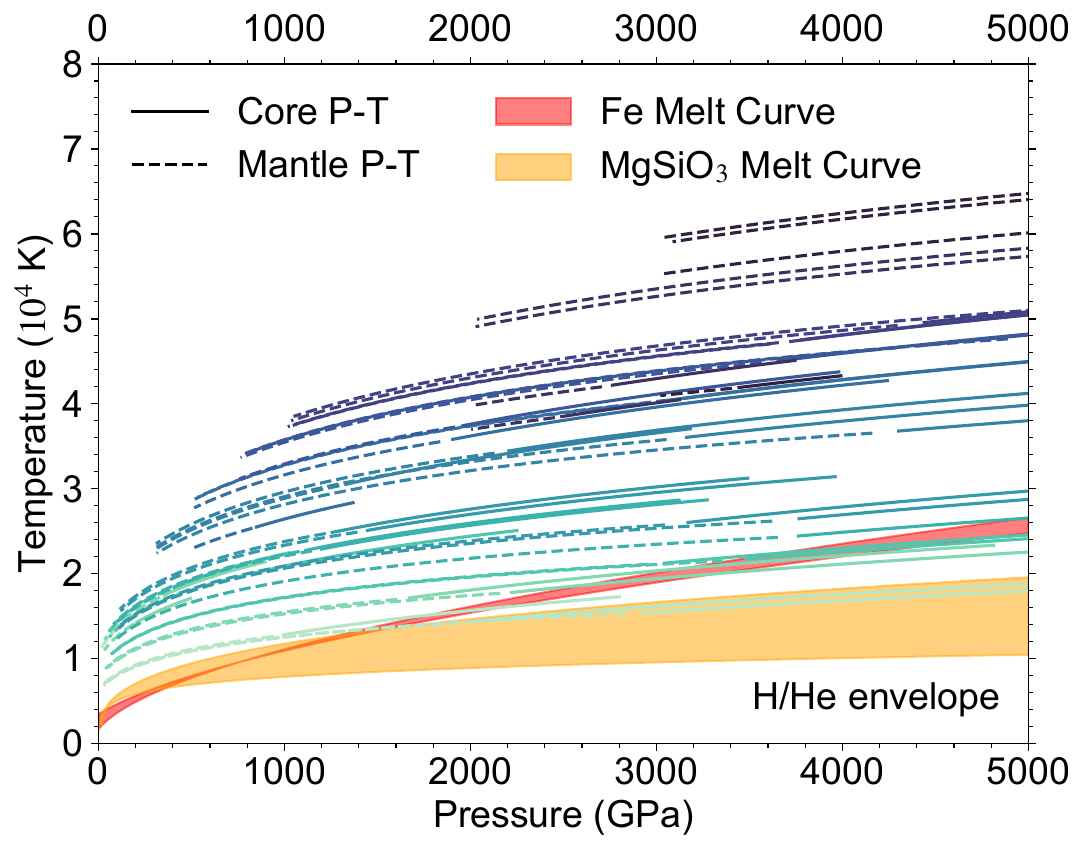}
    \caption{
    P-T profiles of iron core (solid curves) and silicate mantle (dashed curves) below H/He envelopes. 
    For clarity, only $\sim50$ randomly chosen models are shown. Darker color implies higher \peq\ (see Figure \ref{fig:MMF_hhe} for color scheme). The melt curves of iron (red) and MgSiO$_3$ (orange) are plotted for comparison, where the shaded regions represent uncertainties in melting temperatures. 
    Cores below H/He envelopes are generally much hotter than iron and MgSiO$_3$ melt curves, explaining their high MMFs (Figure \ref{fig:MMF_hhe}).}
    \label{fig:hhe_pt_vs_meltCurve}
\end{figure}

\section{Results \& Discussion} \label{sec:results}

\begin{figure*}
    \centering
    \includegraphics[width=\linewidth]{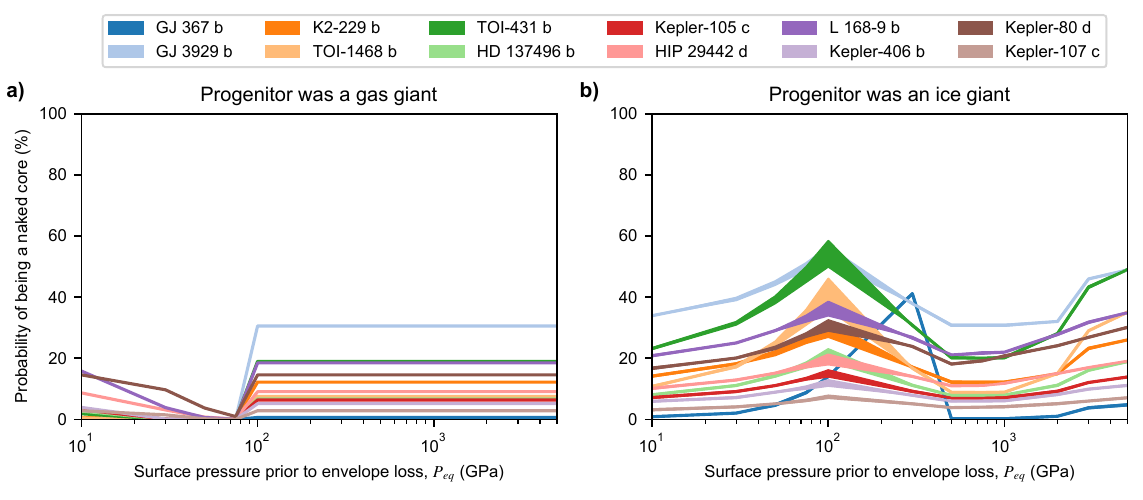}
    \caption{
    The probability that a certain high-density exoplanet is a naked core, as a function of \peq\, assuming that the progenitor was \textbf{a)} a gas giant or \textbf{b)} an ice giant, color-coded by planet.
    In b), the shaded regions are bound between the curves corresponding to coolest and hottest temperature profiles allowed for ice giants. 
    The highest probabilities ($<$ 60\%) are recorded for naked cores with ice giant progenitors around 100 GPa. Most high-density exoplanets have low probabilities to be remnant cores of giant planets.}
    \label{fig:P_naked}
\end{figure*}

\subsection{Naked Cores with Gas Giant Progenitors Are Indistinguishable from Self-compressed Planets}

We find that the naked core hypothesis is highly unlikely to explain the origin of high-density planets if they had gas giant progenitors whose envelope photoevaporated through energy-limited escape. This is because all iron-silicate cores with thick ($\geq10$ GPa) H/He envelopes are fully or almost fully molten (Figure \ref{fig:MMF_hhe}a). 
Because the melting temperature of iron increases more rapidly with pressure than MgSiO$_3$, some naked cores have partially or fully solid iron cores (Figure \ref{fig:MMF_hhe}b). However, such (partially) solid iron cores can only exist if the envelopes are not very massive (\peq\ $\leq100$ GPa). For gas giant progenitors with \peq\ $\geq300$ GPa, temperature effect due to thermal blanketing of thick atmosphere prevails over pressure effects, yielding fully molten iron cores. 

MMF results presented here are robust against uncertainties in planetary interior P-T profiles and melting temperatures. Temperatures inside H/He envelopes are generally much hotter than melting temperatures of iron and silicate (Figure \ref{fig:hhe_pt_vs_meltCurve}). Even though our fully adiabatic P-T model may underestimate interior temperature by $\sim1000$ K, because nonadiabatic effects near core-mantle boundary are ignored (see Section \ref{sec:methods}), such effects are considerably smaller than the temperature gap between P-T profiles of extremely hot cores and melt curves.

Population-wise, we can reject the hypothesis that all high-density planets are remnant naked cores of gas giants through the Kolmogorov–Smirnov (KS) test (Appendix \ref{sec:appendix_probability_HD}) because the populations of naked cores of gas giants have p-value $\ll$ 0.05 (exactly, $<3\times 10^{-27}$) to match that of the observed high-density exoplanets. Consistently, on a planet-to-planet basis, we find that no high-density exoplanet has probability higher than 20\% to be the core of a gas giant, except for GJ 3929 b (probability of at most $\sim30\%$, Figure \ref{fig:P_naked}a). The latter can be explained with GJ 3929 b having the lowest bulk density among high-density exoplanets, and thus the most consistent with a planet of terrestrial (CMF $\sim33$\%) composition within uncertainties. For $P_{\rm eq}>$ 100 GPa, the probability is constant because naked cores achieve 100\% MMF and no fossil compression, so that their M-R relations are indistinguishable from that of a planet with an Earth-like interior structure.

Our finding that the mantle is fully molten for almost all \peq\ suggests that global MO is long-lived. The global MOs do not begin to solidify until \peq\ decreases to $\sim10$ GPa. Even when only a thin (\peq\ $=10$ GPa) envelope remains, $\gtrsim65\%$ of the mantle is still molten. Eroding massive envelopes requires a long time, allowing extensive global MO to stay molten for a timescale on the order of Gyr \citep[see also][]{vazan_contribution_2018}. The longevity of global MO guarantees that the core has sufficient time to rebound as pressure unloads.

The prevalence of thick MOs on planets with substantial H/He envelopes is consistent with previous studies. In sub-Neptunes with compositionally Earth-like cores and H/He atmospheres constituting several percent of the planet's mass, temperature at the magma-atmosphere interface is predicted to be hotter than the silicate liquidus \citep{kite_atmosphere_2020}. Pressure at the magma-atmosphere boundary in such a sub-Neptune is roughly 1--10 GPa, coinciding with the lower limit of \peq\ we consider here. Using an adiabatic temperature model for the interior of sub-Neptunes, \cite{kite_atmosphere_2020} found that MO mass on a $5\,M_\oplus$ core can easily exceed $2\,M_\oplus$ (i.e., MMF $=0.4$) given a $\gtrsim1000$ K equilibrium temperature, similar to the equilibrium temperatures derived by our model.

The longevity of MO on planets with substantial H/He envelope is also well established. \cite{vazan_contribution_2018} reported that sub-Neptunes with H/He envelope mass fractions up to 10\% can stably sustain massive MO on the timescale of several Gyr. Long-lived MO in contact with the envelope suggests active chemical interactions between the core and the atmosphere \citep[e.g.,][]{vazan_contribution_2018, kite_atmosphere_2020, schlichting_chemical_2022, misener_atmospheres_2023}, which may enrich the envelope with metals and the core with volatiles. Such a material exchange may reduce bulk densities of naked cores due to light element inclusion in the core. This density reduction does not impact the robustness of our results, because we aim to place an upper limit on densities of naked cores.

\begin{figure*}
    \centering
    \includegraphics[width=\linewidth]{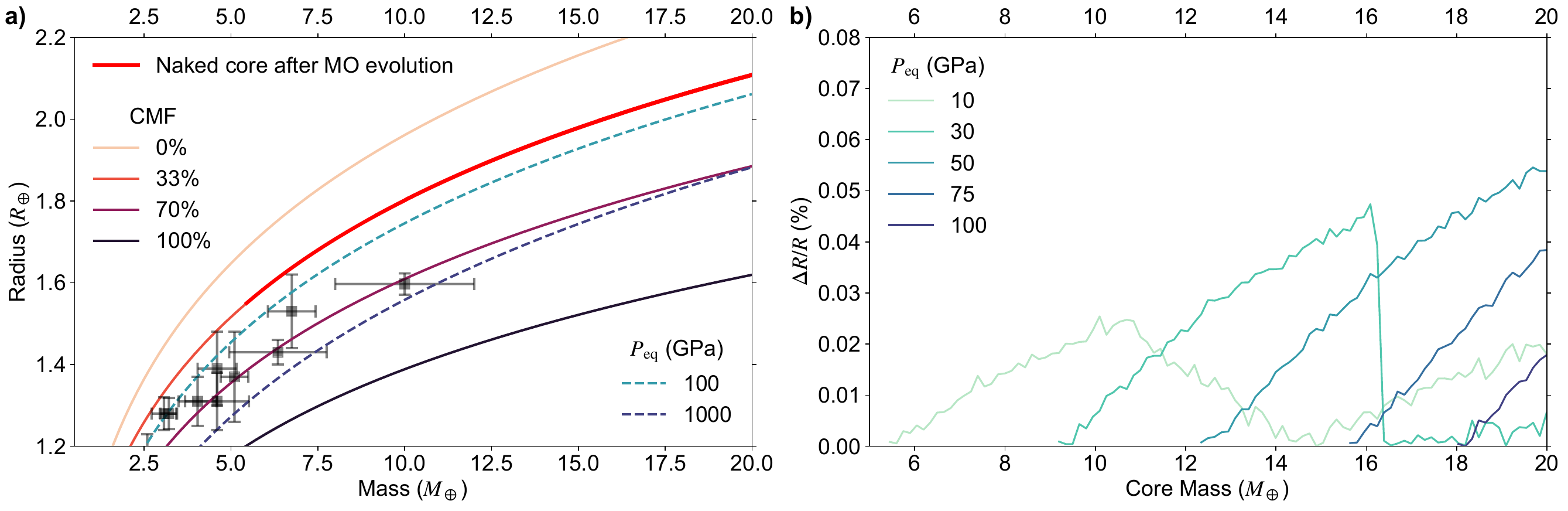}
    \caption{\textbf{a)} Mass-radius relation of naked cores with gas giant progenitors. After MO and envelope evolution, the M-R relation of naked cores (thick red curve) are visually indistinguishable from self-compressed planets with the same 33\% CMF (thin red curve). Self-compressed M-R curves with 0\%, 70\%, and 100\% CMF (solid), as well as fully solid and fossil-compressed M-R curves with \peq\ $=100$ and 1000 GPa (dashed) are plotted for comparison. Squares with error bars represent observed high-density planets (Table \ref{tab:planet_params}). \textbf{b)} Relative radius difference, $\Delta R/R$ in per cent, between naked core M-R curves and the self-compressed M-R curve with CMF $=33\%$. The maximum radius deviation is only $\approx$ 0.06\%. Bulk densities of naked cores evolved from gas giant progenitors are indistinguishable from self-compressed planets.}
    \label{fig:hhe_mr_mo}
\end{figure*}

Due to high MMFs (Figure \ref{fig:MMF_hhe}), after envelope loss, bulk densities of naked cores with gas giant progenitors become indistinguishable from normal rocky planets that never experience external compression by massive envelopes. To validate this claim, we rerun our planetary interior model for all naked cores, taking mantle and iron melting into consideration and generate a new set of M-R curves (Figure \ref{fig:hhe_mr_mo}a). M-R curves for all models are visually indistinguishable from each other or the self-compressed 33\% CMF curve, suggesting very similar bulk densities.

To examine effects due to compression more closely, we calculate the radius difference between naked cores and self-compressed planets, and found excellent agreement with $\Delta R < 0.002\,R_\oplus$ for all models, or a relative difference of $\Delta R/R \leq 0.06\%$ (Figure \ref{fig:hhe_mr_mo}b). Such a radius deviation is orders of magnitudes smaller than typical exoplanet radius measurement uncertainty, where $\Delta R/R < 5\%$ is considered ``precise'' \citep[see e.g.,][]{luque_density_2022, parc_super-earths_2024}. This supports our claim that the naked core hypothesis is very unlikely to explain the origin of high-density exoplanets, which have statistically higher bulk densities than other rocky planets (Table \ref{tab:planet_params} and Figure \ref{fig:P_naked}).

\subsection{Naked Cores with Ice Giant Progenitors Exhibit Small but Non-negligible Density Increase}
For iron-silicate cores shrouded by H$_2$O envelopes that experienced energy-limited escape, we find that radius decrease from self-compressed planets is non-negligible (maximum $\Delta R/R \approx 7\%$), but generally insufficient to explain the origin of high-density planets. Here we elucidate our result in three steps. In Section \ref{sec:h2o_bifurcation}, we explain how phase transitions in H$_2$O envelopes introduce degeneracy in MMF. Section \ref{sec:h2o_MMF} discusses the computed iron core and silicate mantle MMF of naked cores with ice giant progenitors. Finally, in Section \ref{sec:h2o_mr}, we quantitatively discuss how compression by a water layer alters the M-R relation of naked cores compared to self-compressed planets.

\subsubsection{Adiabatic P-T Profiles of H$_2$O Envelopes Are Degenerate} \label{sec:h2o_bifurcation}
For naked core progenitors with H$_2$O envelopes, multiple adiabatic P-T profiles with different central temperatures, $T_c$, can all produce the same surface temperature, $T_{\rm surf}$. This degeneracy in P-T profiles give rise to MMF degeneracy.

\begin{figure}
    \centering
    \includegraphics[width=\linewidth]{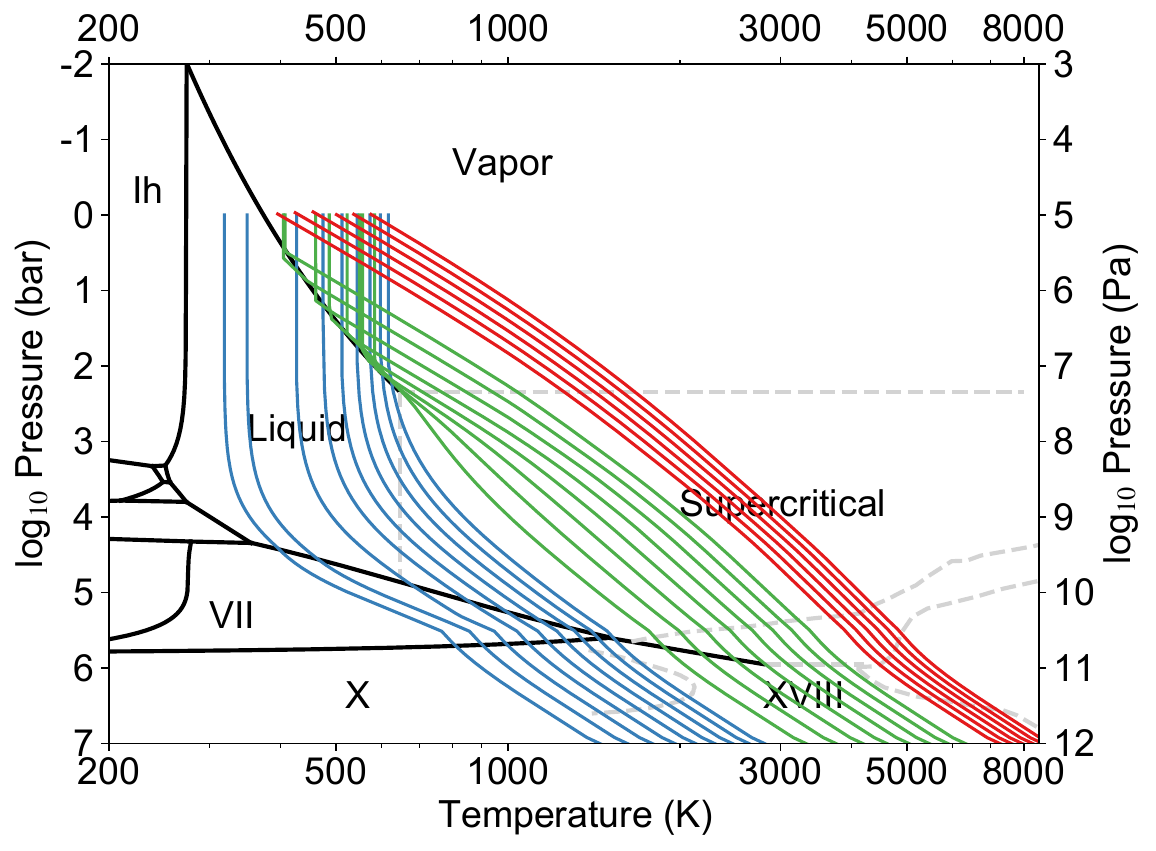}
    \caption{
    H$_2$O phase diagram and P-T profiles with different central temperatures, $T_c$. 
    P-T profiles with a wide range of $T_c$ can reach the same $T_{\rm surf}$ via two paths, namely (blue) the low temperature icy path and (red) the high temperature fluid path. P-T profiles with intermediate $T_c$ (green) can reach similar $T_{\rm surf}$, but their behaviors near the liquid-vapor phase boundary require special treatment (see text).
    P-T profiles in H$_2$O envelopes are degenerate, leading to MMF degeneracy.
    }
    \label{fig:h2o_pt_phase}
\end{figure}

This P-T degeneracy exists because H$_2$O has multiple phases, each having distinct adiabatic gradient (Figure \ref{fig:h2o_pt_phase}). For a moderate $T_{\rm eq}\sim500$ K, there are two viable P-T paths that can make $T_{\rm surf}$ and $T_{\rm eq}$ equal. The first path with lower $T_c$ (Figure \ref{fig:h2o_pt_phase}, blue curves) suggests an icy interior, where the P-T profile traverses various high-pressure ice phases before entering the vapor envelope via a liquid water ocean. Note that for this path, temperature profile of the vapor atmosphere is assumed to be isothermal \citep[see also][]{nixon_how_2021}. The second path with higher $T_c$ (Figure \ref{fig:h2o_pt_phase}, red curves) traverses hot fluid phases (plasma, ionic fluid, and the supercritical) before reaching the steam atmosphere. Between these two paths exist profiles with intermediate $T_c$ that can reach the same $T_{\rm surf}$ (Figure \ref{fig:h2o_pt_phase}, green curves). We force these P-T profiles to be isothermal upon contacting the liquid-vapor phase boundary, to avoid unphysical situations where liquid layers locate above vapor layers. This P-T degeneracy was also observed in models for ``water worlds'' \citep{nixon_how_2021}.

\begin{figure*}[t]
    \centering
    \includegraphics[width=\linewidth]{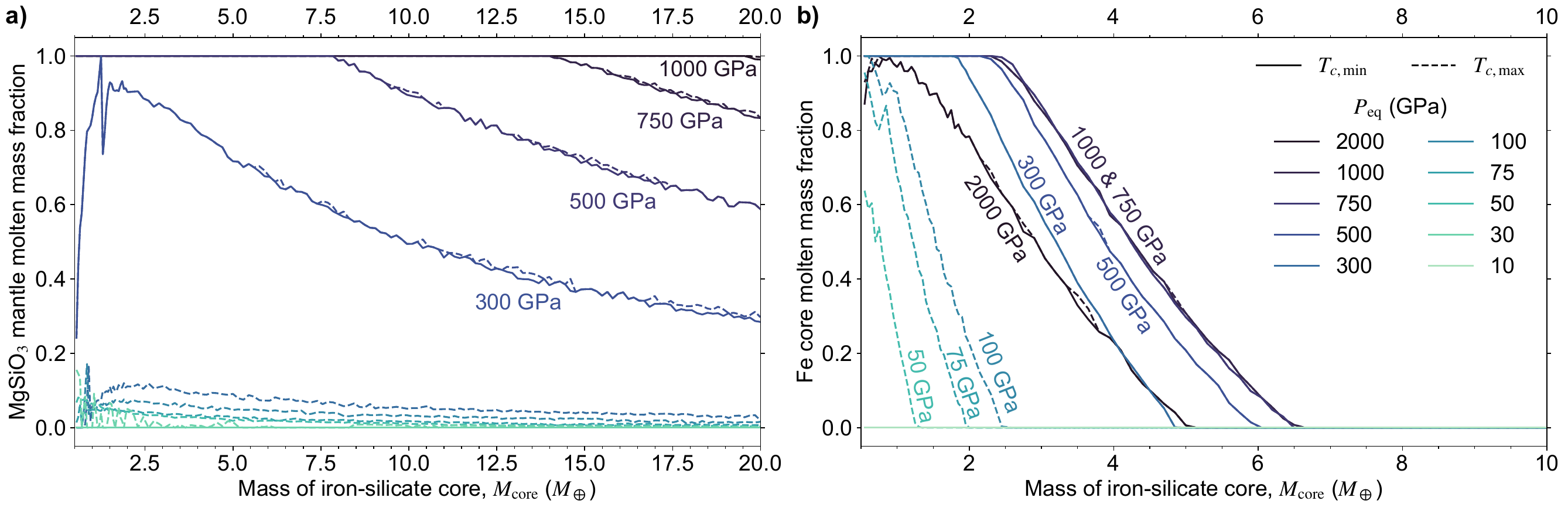}
    \caption{Molten mass fractions of \textbf{a)} silicate mantle and \textbf{b)} iron core for naked cores surrounded by H$_2$O envelopes, as a function of the core mass, $M_{\rm core}$. MMF oscillations reflect uncertainties in $T_c$ solutions introduced by complicated phase transitions of H$_2$O (Figure \ref{fig:h2o_pt_phase}). Iron MMF shows a general decreasing trend with $M_{\rm core}$. Models with \peq\ $>1000$ GPa have fully molten mantles and models with \peq\ $>2000$ GPa have fully solid cores, so these cases are not shown for clarity.}
    \label{fig:MMF_h2o}
\end{figure*}

\begin{figure}
    \centering
    \includegraphics[width=\linewidth]{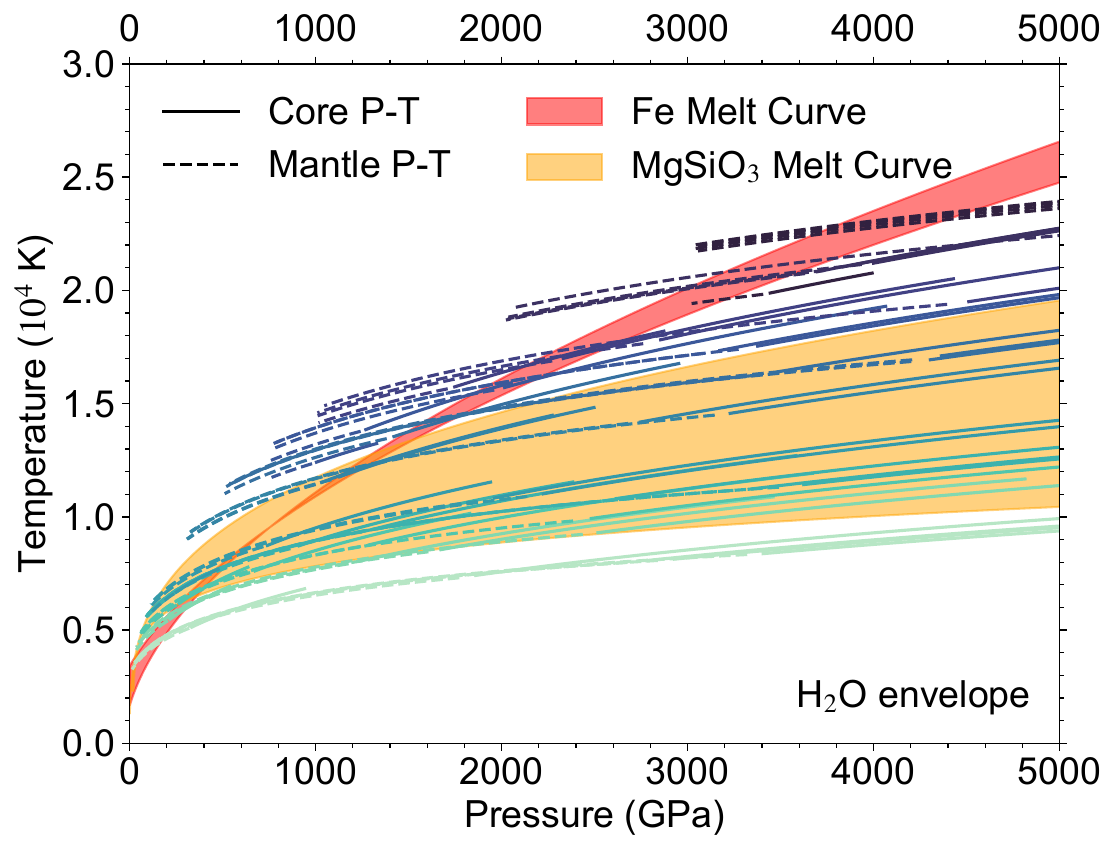}
    \caption{
    P-T profiles of iron core (solid curves) and silicate mantles (dashed curves) below H$_2$O envelopes. 
    For clarity, only $\sim50$ randomly chosen models are shown. Darker color implies higher \peq\ (see Figure \ref{fig:MMF_h2o} for color scheme). The melt curves of iron (red) and MgSiO$_3$ (orange) are plotted for comparison, where the shaded region represents uncertainties in melting temperatures. 
    M-R curves of naked cores with ice giant progenitors are more sensitive to uncertainties in MgSiO$_3$ melt curve, because core P-T profiles and the uncertainty region of the melt curve largely overlap.
    }
    \label{fig:h2o_pt_vs_meltCurve}
\end{figure}

As a result of degenerate P-T profiles, iron-silicate MMFs below H$_2$O envelopes are subject to degeneracy. To address this degeneracy, we report two MMF values for each model. The higher MMF corresponds to the hottest possible interior with maximum central temperature ($T_{c, \rm max}$), while the lower MMF corresponds to coolest possible interior with minimum central temperature ($T_{c, \rm min}$). Valid $T_c$ solutions are searched by the following method: for each synthetic iron-silicate core with H$_2$O envelope, we run the forward planetary interior model with an array of 1500 evenly spaced central temperatures between 800 K and 40,000 K. We search for where $T_{\rm surf} - T_{\rm eq}$ as a function of $T_c$ first and last crosses zero, and use interpolation to find $T_{c, \rm min}$ and $T_{c, \rm max}$ that produce the desired $T_{\rm surf}$.

Note that, however, P-T profiles of H$_2$O envelopes are subject to a few complications. First, the water layer is not necessarily fully adiabatic, analytical \citep{guillot_radiative_2010} or self-consistent \citep{piette_temperature_2020} temperature profiles may be more realistic. Second, massive water-rich planets may accrete substantial amounts of H/He, strengthening the thermal blanketing effect compared to a pure steam greenhouse \citep{koll_hot_2019, innes_runaway_2023}.


\subsubsection{Magma Oceans below H$_2$O Envelopes Are Less Extensive than below H/He Envelopes} \label{sec:h2o_MMF}
We find that iron-silicate cores below H$_2$O envelopes are more likely to stay fossil-compressed and solid, compared to cores below H/He envelopes. Depending on \peq\ and $M_{\rm core}$, cores surrounded by H$_2$O envelopes can be fully molten, partially molten, or fully solid (Figure \ref{fig:MMF_h2o}).

Generally, for naked core progenitors with H$_2$O envelopes, higher \peq\ and lower $M_{\rm core}$ imply more extensive mantle melting (Figure \ref{fig:MMF_h2o}a). When the envelope is very massive (\peq\ $\geq1000$ GPa), complete or almost complete mantle melting is expected for most models, with the only exception being \peq\ $=1000$ GPa models with the most massive cores. When envelope mass is intermediate ($300\leq$ \peq\ $\leq750$ GPa), partial to total mantle melting is expected. When envelope mass is low (\peq\ $\leq100$ GPa), the mantle is expected to be completely or nearly completely solid: MgSiO$_3$ mantle MMF does not exceed 0.2.

For the iron core, pressure effects on MMF are more significant than the silicate mantle (Figure \ref{fig:MMF_h2o}b), because the melting temperature of iron increases more rapidly with pressure compared to MgSiO$_3$ \citep[e.g.,][]{kraus_measuring_2022, gonzalez-cataldo_ab_2023}. 
When the envelope is massive (\peq\ $\geq1000$ GPa), iron core MMF for a given core mass drops as \peq\ increases, reaching zero for all models with \peq\ $\geq3000$ GPa. When the envelope mass is intermediate ($50\leq$ \peq\ $\leq750$ GPa), iron core MMF increases with increasing \peq\ and partial melting is expected for small ($\lesssim 5\,M_\oplus$) $M_{\rm core}$. When the envelope is thin (\peq\ $\leq30$ GPa), temperature in the planet's interior is insufficient to melt the iron core, leading to zero MMF. When the iron-silicate core is massive ($M_{\rm core} \gtrsim 7\,M_\oplus$), iron core MMF becomes zero for all models, because the pressure effects on melting temperature prevails inside massive cores.

MMF results reported here are more sensitive to planetary interior temperature and melting temperature uncertainties. Unlike cores surrounded by H/He envelopes (Figure \ref{fig:hhe_pt_vs_meltCurve}), cores surrounded by H$_2$O envelopes are generally cooler, with P-T profiles overlapping with uncertainty regions of iron and silicate melting temperatures (Figure \ref{fig:h2o_pt_vs_meltCurve}). We mitigate this problem by choosing the hottest melt curves, so the bulk densities of naked cores after MO evolution we report are robust upper bounds.


\subsubsection{Naked Cores with Ice Giant Progenitors Are Less Dense than High-Density Planets} \label{sec:h2o_mr}

\begin{figure}
    \centering
    \includegraphics[width=\linewidth]{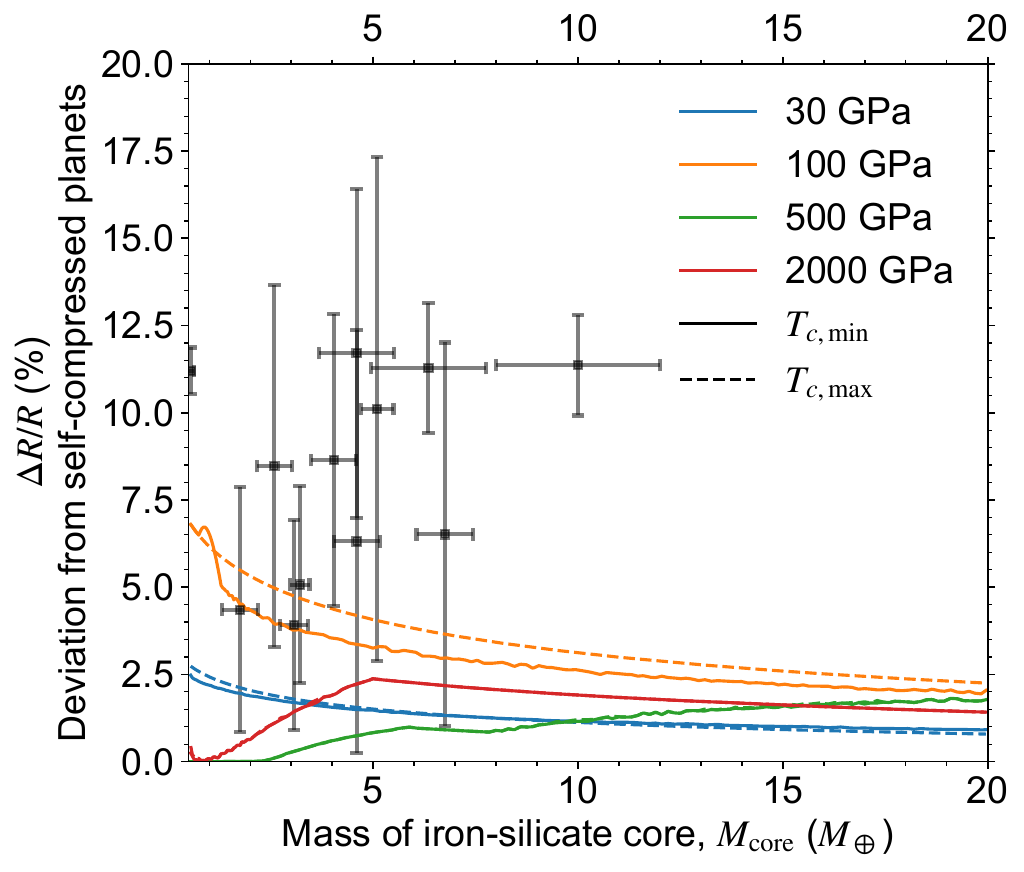}
    \caption{Relative radius deviation, $\Delta R/R$ (\%), between naked cores with ice giant progenitors after MO evolution and self-compressed planets with the same 33\% CMF. $\Delta R/R$ deviations between high-density exoplanets (Table \ref{tab:planet_params}) and the self-compressed 33\% CMF M-R curve are shown as solid circles with error bars. Only four \peq\ cases are shown for clarity. The curves do not substantially overlap with uncertainty ellipses of high-density planets, implying discrepancy between the naked core hypothesis and observation.}
    \label{fig:h2o_mr_curves}
\end{figure}

At least some naked cores that evolved from ice giant progenitors have partially, if not fully, fossil-compressed and solid interiors, as suggested by the silicate mantle and iron core MMFs (Figure \ref{fig:MMF_h2o}), implying high bulk densities. High bulk densities would produce a radius decrease of $\Delta R/R < 7\%$ compared to self-compressed planets with the same 33\% CMF (Figure \ref{fig:h2o_mr_curves}). The exact radius decrease is dependent on \peq\ and $M_{\rm core}$.  

For naked cores whose progenitors have thin (\peq\ $=30$ GPa) envelopes, both iron MMF and silicate MMF are zero or close to zero (Figure \ref{fig:MMF_h2o}). Fully solid and fossil-compressed naked cores produce smoothly decreasing $\Delta R/R$ from $\sim3\%$ to $\sim1\%$.
For the \peq\ $=100$ GPa models, the largest discrepancy is seen between $T_{c, \rm min}$ and $T_{c, \rm max}$ models, implying that the H$_2$O phase effects on P-T profiles (Figure \ref{fig:h2o_pt_phase}) is most significant in this pressure range. $\Delta R/R$ deviations from self-compressed planets are also largest for this \peq, ranging from $\approx7\%$ for the least massive cores to $\approx2\%$ for the most massive cores.
For the \peq\ $=500$ GPa models, $\Delta R/R$ increases with $M_{\rm core}$, because mantle MMF gradually increases with $M_{\rm core}$ too (Figure \ref{fig:MMF_h2o}a), producing a maximum deviation of $\approx2\%$. 
Finally, for models with \peq\ $=2000$ GPa, because the mantle is always fully molten and the core is fully solid for $M_{\rm core} \gtrsim 5\,M_\oplus$, we observe a $\Delta R/R$ curve with two segments. Maximum radius deviation is reached around $5\,M_\oplus$ at around 2\%.

In summary, fossil-compressed solid parts of naked cores with ice giant progenitors decrease their radii by at most 7\%, producing M-R curves between the self-compressed CMF $=33\%$ and CMF $=50\%$ curves. 

The upper limit of $\Delta R/R$ ($\approx7\%$) exceeds the observational exoplanet radius uncertainty limit ($\approx5\%$) and is therefore non-negligible. Motivated by this, we study whether such a compression is sufficient to match the sizes of high-density exoplanets as described in Appendix \ref{sec:appendix_probability_HD}. 
Population-wise, we find that we can reject the 
hypothesis that all high-density planets are naked cores of ice giants through the KS test (Appendix \ref{sec:appendix_probability_HD}) because the populations of naked cores of gas giants have p-value $\ll$ 0.05 (exactly, $<8\times10^{-12}$) to match that of the observed high-density exoplanets. Consistently, we find that no high-density exoplanet has probability higher than 50--60\% to be the core of an ice giant (Figure \ref{fig:P_naked}b). Interestingly, we observe that the  peak in the probability plot in Figure \ref{fig:P_naked}b at $P_{\rm eq}=100$ GPa for all high-density exoplanets except the smallest of them, GJ 367 b, whose probability peaks at higher pressures ($P_{\rm eq}=300$ GPa). We explain this peak as follows. As $P_{\rm eq}$ increases, not only compression and bulk density increases, but also temperature. When $P_{\rm eq}$ becomes sufficiently high that the internal temperature overshoots the melting temperature of the materials, bulk density starts decreasing because MMF increases. This leads to the existence of a point of optimum, which we record to be $P_{\rm eq}\sim$ 100 GPa for planets with radii of 1--2 $R_\oplus$ and higher for smaller planets. The probability of being a naked core increases again for $P_{\rm eq} > 2000$ GPa because the iron core fully solidifies under enormous pressure (Figure \ref{fig:MMF_h2o}b), increasing the predicted bulk density.

We conclude that most high-density planets are unlikely to be naked cores that once had H$_2$O envelopes. Note that, however, this claim is subject to some uncertainties introduced by, for example, temperature profile assumed for the H$_2$O layer and valid under the assumption of energy-limited envelope escape.

\section{Conclusions} \label{sec:dis_conclusion}

We use a planetary interior model combined with an adiabatic temperature profile to show that the naked core hypothesis is unlikely to explain the origin of high-density planets. This is because regardless of whether the envelopes on naked core progenitors are composed of H/He or H$_2$O, a large fraction, if not all, of the iron-silicate cores would be molten at some point during the envelope evolution. Due to the much lower viscosity of magma compared to solid rock and iron, the presence of thick global magma ocean implies that naked cores can rapidly rebound to a self-compressed state as the envelope is eroded. 

When taking magma ocean evolution into consideration, naked cores have densities that are hardly distinguishable from self-compressed planets if they used to have H/He envelopes. The maximum radius deviation is $\Delta R/R < 0.06\%$ (Figure \ref{fig:hhe_mr_mo}). This means that the remnant naked cores of giant planets are observationally indistinguishable from self-compressed planets with Earth-like composition that never experience the evolution shown in Figure \ref{fig:flowchart}, if the only known information is bulk planetary density. If H$_2$O envelope is assumed instead, radii of naked cores are smaller than self-compressed planets by a maximum deviation of $\Delta R/R < 7\%$, owing to a lower molten mass fraction (Figure \ref{fig:h2o_mr_curves}). 


Under the assumption of energy-limited atmospheric escape, our finding that the naked cores of giant planets should be partially-to-fully molten invalidates one of the basic assumptions of the theory of high-density naked cores as formulated by \citet{2014RSPTAMocquet}, that is, that the naked cores are fully solid iron-silicate cores (mass-radius relation in Figure \ref{fig:mr_combined}, bottom). By factoring in the effect of decompression due to magma ocean development, the corresponding population of naked cores have Kolmogov-Smirnoff probability $\ll$ 0.05 to be drawn from the same population of planets as the high-density exoplanets, regardless of whether the progenitors of the former are gas giants or ice giants. On a planet-by-planet basis, we find that the high-density exoplanets have overall probability $<30\%$ and $<20\%$ across all values of confining pressure, $P_{\rm eq}$ for the ice giant and gas giant case, respectively. The two exceptions are the two high-density exoplanets with the lowest density: GJ 3929 b and TOI 431 b, whose probability to be the naked core is lower than $\sim$ 50\%. Importantly, the probability that GJ 3929 b is a naked core of an ice giant is similar to that of formation by a mantle-stripping giant impact as estimated by \citet{2024Cambioni_inprep}, so we cannot distinguish between the two hypothesis.

These findings imply that most of the high-density exoplanet we see today are not naked cores of giant planets. The two remaining hypothesis are preferential accretion of metal-rich refractory materials \citep{2020ApJAguichine, 2022AAJohansen}, or mantle-stripping giant impact \citep{2009ApJMarcus, 2022MNRASReinhardt}, with the former being favored based on planet formation studies \citep{2024Cambioni_inprep}. Note that our results do not rule out the naked core evolution outlined in Figure \ref{fig:flowchart}. It still physically possible that some of the rocky planets we see today used to reside under thick H/He or H$_2$O envelopes, but with probabilities generally lower than 20--30\%.

Finally, by considering a fossil-compressed core within a giant planet (Figure \ref{fig:flowchart}), we implicitly assume a concentrated core geometry. However, it is increasingly accepted that Jupiter and Saturn have fuzzy cores \citep[e.g.,][]{stevenson_jupiters_2020, mankovich_diffuse_2021, nettelmann_theory_2021}. For ice giants, the idea that rock and ice are not separate but miscible to create compositional gradients is also becoming increasingly popular \citep[e.g.,][]{helled_interiors_2020, 2020RSPTATeanby, vazan_new_2022}. A fuzzy core geometry renders the naked core hypothesis even less likely by naturally reducing the fraction of the core that can stay solid and fossil-compressed, if such a solid core exists at all. \\
\\
\noindent
ZL acknowledges funding from the Center for Matter at Atomic Pressures (CMAP), a National Science Foundation (NSF) Physics Frontiers Center, under Award PHY-2020249. SC acknowledges funding from the Crosby Distinguished Postdoctoral Fellowship of the Department of Earth, Atmospheric and Planetary Sciences, Massachusetts Institute of Technology. The authors acknowledge the MIT SuperCloud and Lincoln Laboratory Supercomputing Center for providing high performance computing resources that have contributed to the research results reported within this paper.

%

\vspace{5mm}
\facilities{Exoplanet Archive \citep{nasa_exoplanet_science_institute_planetary_2020}}


\software{\texttt{Matplotlib} \citep{Hunter_2007_matplotlib}, \texttt{NumPy} \citep{harris2020array}, \texttt{SciPy} \citep{2020SciPy-NMeth}}



\appendix

\section{Planetary Interior Model Validation} \label{sec:appendix_validation}

Here, we thoroughly validate our planetary interior model by comparing it to relevant models in the literature. We first generate Earth models to test the model's robustness for simulating rocky bodies. Then, we run interior models for sub-Neptune K2-18 b to test its performance for H$_2$O- and H/He-rich bodies. 

Mass and radius calculated for an Earth model by our planetary interior model are $0.9996\,M_\oplus$ and $0.9708\,R_\oplus$, which only deviate from the ground truth by 0.04\% and 2.92\%, respectively, assuming 33\% CMF and an isothermal interior. The calculated mass and radius deviations are smaller than typical exoplanet M-R uncertainties, where $\Delta M/M < 25\%$ and $\Delta R/R < 5\%$ are considered ``precise'' \citep[see e.g.,][]{luque_density_2022}. Note that because our model does not include details such as light elements in core and mantle phase transitions, a perfect match in mass and radius is not expected.

To validate the adiabatic thermal model for iron and silicates, we generate a P-T profile for Earth using Equation (\ref{eq:boujibar_ad_t}) and compare it to the geotherm \citep{hirose_composition_2013}. The adiabatic models are in good agreement with the geotherm in the core, with a $\lesssim2\%$ temperature deviation. Our fully adiabatic P-T model, however, fails to capture the conductive layer near the core-mantle boundary \citep[e.g.,][]{valencia_internal_2006, wagner_rocky_2012}, leading to a P-T profile that is $\sim1000$ K hotter than the geotherm in the mantle.

To validate our planetary interior model for H$_2$O- and H/He-rich planets, we generate three K2-18 b models with significantly different compositions \citep[following][]{madhusudhan_interior_2020}. The first model has an extremely iron-rich core and thick H/He envelope ($x_{\rm Fe} = 94.7\%$, $x_{\rm H_2O} = 0.3\%$, and $x_{\rm H/He} = 5\%$), the second model is a sub-Neptune ($x_{\rm Fe} = 14.85\%$, $x_{\rm MgSiO_3} = 30.15\%$, $x_{\rm H_2O} = 54.97\%$, and $x_{\rm H/He} = 0.03\%$), and the third model assumes a ``water world" composition ($x_{\rm Fe} = 3.3\%$, $x_{\rm MgSiO_3} = 6.7\%$, $x_{\rm H_2O} = 89.994\%$, and $x_{\rm H/He} = 0.006\%$). $x_i$ here represents the mass fraction of material $i$. Masses and radii predicted by our model deviate from the observed values ($M_p = 8.63\pm1.35\,M_\oplus$, $R_p = 2.610\pm0.087\,R_\oplus$; \citealt{Benneke_water_2019}) by $\Delta M/M = 0.16\%$ and $\Delta R/R = 5.51\%$, $\Delta M/M = 0.01\%$ and $\Delta R/R = 1.45\%$, and $\Delta M/M < 10^{-7}$ and $\Delta R/R = 2.72\%$, for the three models, respectively. These mass and radius deviations are smaller than typical exoplanet mass and radius uncertainties \citep[e.g.,][]{luque_density_2022}, except for the first model with thick H/He envelope, because of the sensitive dependence of H/He envelope radius on internal temperature and mean molecular weight.

\section{Modeling Envelope Evaporation} \label{sec:appendix_evaporation}
Here, we discuss the equation we use to calculate the timescale of envelope photoevaporation in detail. The photoevaporation timescale, $\tau_{\rm photo}$, is defined as \citep{owen_evaporation_2017}
\begin{equation} \label{eq:tau_photo}
\begin{split}
    \tau_{\rm photo} &= 210 \text{ Myr } \left(\frac{\epsilon}{0.1}\right)^{-1} \left(\frac{L_{\rm HE}}{10^{-3.5} L_\odot}\right)^{-1} \\
    &\times \left(\frac{P}{10 \text{ days}}\right)^{1.41} \left(\frac{M_*}{M_\odot}\right)^{0.52} \left(\frac{f}{1.2}\right)^{-3} \\
    &\times \left(\frac{\tau_{\rm KH}}{100 \text{ Myr}}\right)^{0.37} \left(\frac{\rho_c}{5.5 \text{ g cm}^{-3}}\right)^{0.18} \left(\frac{M_c}{5M_\oplus}\right)^{1.42} \\
    &\times \begin{cases}
        \left(\frac{\Delta R}{R_c}\right)^{1.57} & \text{ if } \Delta R / R_c < 1 \\
        \left(\frac{\Delta R}{R_c}\right)^{-1.69} & \text{ if } \Delta R / R_c \geq 1,
    \end{cases}
\end{split}
\end{equation}
where $\epsilon$ is the dimensionless efficiency factor, which is assumed to be 0.1 following \cite{owen_evaporation_2017}. $L_{\rm HE}$ is the high-energy luminosity of the host star. We make the simplifying assumption that $L_{\rm HE}$ and the bolometric luminosity of the star, $L_{\rm bol}$, are related by $L_{\rm HE} = 10^{-3.5}L_{\rm bol}$. $P$ is orbital period of the planet. $M_*$ is stellar mass. $f$ is a constant such that radius of the radiative-convective boundary, $R_{\rm RCB}$, is related to the planetary radius by $R_p = f R_{\rm RCB}$. For simplicity, we assume that $f=1$. $\tau_{\rm KH}$ is the Kelvin-Helmholtz timescale, or the cooling timescale, and is typically assumed to be 100 Myr or the planet's age, whichever is larger \citep{owen_evaporation_2017}. Because ages of high-density planets are typically unknown or subject to large uncertainties (Table \ref{tab:planet_params}), we assume $\tau_{\rm KH}=1$ Gyr, the assumed nominal planet age. $\rho_c$ is the iron-silicate core density that equals 5.5 g cm$^{-3}$ without external compression, because we consider an Earth-like iron-to-silicate ratio. When fossil-compressed, $\rho_c$ is scaled based on the level of compression. $M_c$ is the core mass. $\Delta R/R_c$ is the ratio of envelope thickness to core radius. After applying the above simplifying assumptions, $\tau_{\rm photo}$ depends only on $P$, $\rho_c$, $M_c$, and $\Delta R/R_c$. 

Note that Equation (\ref{eq:tau_photo}) is typically applied to planets with H/He envelopes. When applied to H$_2$O escape, fractionation between oxygen escaping to space and oxygen accumulating in the atmosphere needs to be taken into consideration \citep[e.g.,][]{luger_extreme_2015}. Here we take the limit that energy-limited escape is highly efficient, such that all O escape to space together with H, so that the above expression can be applied to ice giant envelope erosion without modification. Even if this assumption does not hold true, an accumulated oxygen-rich atmosphere keeps the core warmer than an airless body, increasing MMF. Thus, the upper limit on densities of naked cores we place will still be robust.

To calculate adiabatic P-T profiles, we need an outer boundary condition for temperature. This outer boundary condition is obtained by solving Equation (\ref{eq:tau_photo}) for each of the synthetic naked core progenitors to find the right orbital period $P$ to make $\tau_{\rm photo}=1$ Gyr. We adopt this conservative choice for $\tau_{\rm photo}$ so that the equilibrium temperature of naked core progenitors are lower and therefore part of the  core may remain solid throughout evolution. However, note that in reality $\tau_{\rm photo}$ can be smaller, because atmospheric erosion is typically the most efficient within the first 100 Myr, when the young host star has much higher $L_{\rm HE}$ than old stars \citep{owen_evaporation_2017}. We assume that the host star is Sun-like with mass equals $M_\odot$ and luminosity equals $L_\odot$ to solve for the planet's equilibrium temperature $T_{\rm eq}$ while assuming zero albedo for simplicity.
The assumption of zero albedo is justified because hot Jupiters, which are representative of naked core progenitors, generally have low albedos below 0.2 \citep{mallonn_low_2019}. For planets with H/He envelopes, the calculated $T_{\rm eq}$ ranges from $\sim750$ K to $\sim2500$ K, for $P_{\rm eq}$ from 10 to 5000 GPa, respectively. For planets with H$_2$O envelopes, the calculated equilibrium temperatures range from roughly 300 to 1400 K, where planets with higher $P_{\rm eq}$ and higher $M_{\rm core}$ have higher $T_{\rm eq}$.

Once the outer temperature boundary condition, $T_{\rm eq}$, is solved, we use a bisection method to solve for the correct central temperature, $T_c$, that produces a surface temperature that equals $T_{\rm eq}$ within some small tolerance (on the order of 1 K).

\section{Estimating the Likelihood of High-density Exoplanets Being Naked Cores} \label{sec:appendix_probability_HD}

We compare our predictions for the mass-radius relations of naked cores to the measured $M$ and $R$ of high-density exoplanets (Table \ref{tab:planet_params}), to the evaluate the likelihood that high-density exoplanets are naked cores of giant planets. We distinguish between the hypothesis that \textit{all} high-density exoplanets are naked cores, and that \textit{some} high-density exoplanets are naked cores.

We evaluate the hypothesis that \textit{all} high-density exoplanets are naked cores as follows. First, for a given scenario (that is, a naked core of a gas giant or an ice giant with $P_{\rm eq}$), we build a synthetic population of 12,000 naked cores, 1000 for each of the observed high-density exoplanets. Each batch of 1000 naked cores have radii bootstrapped within the error bar of possible radii for a certain high-density exoplanet, and mass defined by the M-R relation from the naked core scenario based on such radii for $P_{\rm eq}$ randomly sampled between 0 and 5000 GPa. Next, we build a synthetic population of high-density exoplanets by bootstrapping 12,000 ($M$,$R$) values within their error ellipses, 1000 for each exoplanets. Finally, we compare the 2-dimensional (2-D) mass-radius distribution of the naked cores to that of the high-density exoplanets by means the 2-D Kolmogorov–Smirnov (KS) test\footnote{Using the \texttt{ndtest} package: \href{https://github.com/syrte/ndtest}{github.com/syrte/ndtest}} to compute the probability of the null hypothesis that the two populations are drawn from the same underlying population. For the KS test, this probability is represented by the p-value, which if $<$0.05 rejects the null hypothesis. We repeat the above procedure 100 times  for each scenario to get a p-values statistics. 

To evaluate the hypothesis that \textit{some} high-density exoplanets are naked cores, we compute the rate at which a given scenario form simulated naked cores with $M$ and $R$ similar to those of the observed high-density exoplanets, on a planet-by-planet basis. First, for a given observed planet, we randomly sample 10$^6$ ($M$,$R$) couples within a 2-D Gaussian centered at its nominal mass and radius and with standard deviations equal to its maximum error in mass and radius. Next, we compute the probability that this planet is explained by a naked core with $P_{\rm eq}$ (either for the gas-giant or ice-giant scenario) as the fraction of the 10$^6$ synthetic M-R pairs that have densities above the naked core density curve.

\bibliography{high_density_MR}{}
\bibliographystyle{aasjournal}



\end{document}